\documentstyle[epsfig,psfig]{article}
\newcommand{\ccbar}{c\bar c}
\newcommand{\jp}{J/\psi}
\newcommand{\nstar}{N^{*}}
\newcommand{\st}[1]{|#1\rangle}

\newcommand{\jpc}{J^{PC}}
\def \qqbar {q\bar q}
\newcommand{\str}[1]{|#1\rangle}
\newcommand{\elmnt}[3]{\langle #1|#2|#3\rangle}

\title { A Study of the Roper Resonance as a Hybrid State from  $J/\psi $ Decays}
\author{R.G. Ping$^{a,b}$\footnote{Corresponding author. E-mail:pingrg@mail.ihep.ac.cn, Fax: 86 10 88233085, Tel:86 10 88236157
(o); 86 10 88233050(h)},H.C.Chiang$^{a,b,c,d}$,B.S.Zou$^{a,b,c,d}$\\
a) CCAST(World Lab.), P.O.Box 8730, Beijing 100080;\\
b) Institute of High Energy Physics,Chinese Academy of Sciences,\\ P.O.Box 918(4), Beijing 100039,P.R.China;\\
c)Institute of Theoretical Physics, Chinese Academy of
Sciences,P.O.Box 2735,P.R.China\\ Beijing
100080, P.R.China\\
d) Center of Theoretical Nuclear Physics, National Laboratory of
Heavy Ion Accelerator, \\ Lanzhou 730000, P.R.China  \\}

\begin{document}
\maketitle
\renewcommand\baselinestretch{1.5}  
\abstract{The structure of the Roper resonance as a hybrid baryon
is investigated through studying the transitional amplitudes in
$\jp \to \bar p\nstar,\bar\nstar\nstar$ decays. We begin with
perturbative QCD  to describe the dynamical process for the
$\jp\to 3\bar q+3q$ decay to the lowest order of $\alpha_s$, and
by extending the modified quark creation model to the $\jp$ energy
region to describe the $\jp\to 3\bar q+3q +g$ process. The
non-perturbative effects are incorporated by a simple quark model
of baryons to evaluate the angular distribution parameters and
decay widths for the processes $\jp\to \bar p\nstar,\bar
\nstar\nstar$. From fitting the decay width of $\jp\to\gamma p\bar
p$ to the experimental data, we extract the quark-pair creation
strength $g_I=15.40$ GeV.  Our numerical results for $\jp\to \bar
p\nstar,\bar\nstar\nstar$ decays show that the branching ratios
for these decays are quite different if the Roper resonance is
assumed to be a common $3q$ state or a pure hybrid state. For
testing its mixing properties, we present a scheme to construct
the Roper wave function by mixing $\st{qqqg}$ state with a normal
$\st{qqq,2s}$ state. Under this picture, the ratios of the decay
widths to that of the $\jp\to p\bar p$ decay are re-evaluated
versus the mixing parameter. A test of the hybrid nature of the
Roper resonance in $\jp$ decays is discussed.
}\\
PACS number(s):  14.20.Gk,13.25.Gv,12.39.Mk,24.85.+p,12.38.Bx

\section{Introduction}
The theoretical studies on the hadronic structure remain as long
outstanding problems in particle physics. About four decades ago,
the naive quark model was proposed to count for SU(3) symmetries
in hadronic spectroscopies\cite{qm}. After that, much progresses
have been made in many theoretical aspects, such as the
predictions on the mass spectrum, decay rate and electromagnetic
coupling properties for hadrons (for details,see recent
review\cite{capstick&r}). So far, much more phenomenological
models based on the quark constituents have been developed with a
large number of successes for further understanding the relations
between quarks' degrees of freedom and fundamental field of
quantum chromodynamics (QCD) theory. Despite these successes the
quark models acquired, they are still confronted with some
challenges. One of the outstanding questions is the fact that, in
addition to the conventional quark states, QCD theory also
predicts the existence of hadrons with 'excited glue', which is
called hybrid states historically. Therefore, searching for these
new hadronic states is of great importance both to QCD theory and
the common quark models. If no hybrid meson/baryon states are
found, it would imply that our current understanding of QCD theory
should be modified, and the dynamics within quark model would have
to be changed. On the other hand, if an unconventional baryon
state is discovered in experiment, it would be necessary to
determine whether it is a hybrid state or a common quark state, or
their mixture. However, the answer to this question seems to be
more difficult from experimental aspects due to the fact that it
depends on a comparison with theoretical predictions on the
specific properties for the new states.

Therefore, the identification of hybrid states would require a
close collaboration between experimental and theoretical
experts\cite{senba}. As our knowledge about common hadronic
structures is mainly acquired from their mass spectrum, hadronic
spectroscopy will continue to be a key tool in this field. Various
models and methods have been used to predict the spectrum of
hybrid mesons/baryons, such as the bag model,QCD sum rule,the flux
tube model,and so on. Though each model assumes a particular
description of excited glue, fortunately they often reach similar
conclusion regarding the quantum numbers and approximate masses of
these states. For instance, the predictions on the light hybrid
mesons are in good agreement with each other, with the so-called
exotic number $J^{PC}=1^{-+}$ and the mass about $1.5\sim 2.0$GeV.
Experimentally we now have some candidates for $\jpc$ exotic
mesons, for example, $\pi_1(1400)$ with $1^{-+}$ seen in $\eta
\pi$\cite{pi1400}, and $\pi_1(1600)$ seen in $\rho\pi,\eta\pi$ and
$b_1\pi$\cite{pi1600}. In baryon sector, the Roper resonance
N*(1440) has been suggested to be a potential candidate of hybrid
baryons for a long time. In the pioneering
 work of the evaluation of the hybrid baryon mass, Barnes and Close once used the
 bag model to predict that the lightest hybrid baryon was an
 "extra" ${1 \over 2}^+N^*(P_{11})$ state with energy at about 1.6 GeV\cite{barnes}.The
 subsequent identification of this hybrid baryon with the Roper
 resonance was confirmed basically by Golowich,Haqq and Karl\cite{golowich}. The
 estimation of its mass was also carried out by using the QCD  sum
 rule, a general prediction came with a conclusion that the lightest ${1 \over 2}^+$
 hybrid baryon is near 2.1 GeV\cite{martynenko}. Recently, this value has been corrected
 by Kisslinger to be about 1.5 GeV, very suggestive of the
 the Roper resonance \cite{kisslinger}.It seems that
 one might find the state with excited gluonic degrees of freedom  by studying baryonic mass spectrum.
 Unfortunately to identify a hybrid baryon by the mass
 spectrum remains an unconquerable difficulty, mainly due to the lack of
 $J^P$ exotics (this situation is quite different from that in hybrid mesons). Hence, for their identification we have
 to predict other experimental observables characterized by
 their distinct properties as a signature of hybrid baryons. For
 example,strong decay amplitudes\cite{Jdecays}, photoproduction
  and electroproduction(EM) amplitudes and so on, can be served for this purpose\cite{li}.
 Especially,the study of the $N^*$
 spectroscopy from the $\jp$ hadronic decays at
 Beijing Electronic-Positron Collider (BEPC) may be an alternative
 tool to identify the nature of the Roper resonance\cite{zou}. One might expect hybrid
 baryons to have a larger production amplitudes from $\jp$ decays for hybrid baryons
 than the conventional $qqq$ baryons, because the three gluonic intermediate
 states produced in $\jp$ annihilations may have a large overlap with
 final hybrid baryons. Furthermore, this approach has an additional advantage
 that it is an isospin I=1/2 filter,so that no $\Delta$ (and hybrid
 $\Delta$)states are present to complicate the analysis.

 Recently, BESII has finished data-taking for 58 million
 more $\jp$ events, which is about two order of magnitude more
 statistics than MARKII  data, and one order of magnitude more
 statistics than BESI data. With such statistics, partial wave
 analysis of relevant channels are possible,
 the angular distributions and the decay widths
  of processes $\jp \to \bar{p}N^*(1440)$ and
 $\jp \to \bar{N}^*(1440)N^*(1440) $  are expected to be available from partial
 wave analysis in the near future.

What properties are able to serve as a signature to distinguish
the Roper resonance between a hybrid baryon and a conventional
three quark baryon? As well-known, the spatial distribution of
constituent components bound in the Roper resonance is described
as the first radially excited state in the naive quark model,
while as a ground state in the hybrid picture. These different
structure pictures , together with their different spin-flavor
structures, may provide us an effective method to test the hybrid
nature of the Roper resonance through studying transitional
amplitude involved in a given $\jp$ decay process. In this paper
we pick out some relevant processes, in which the $\jp$ particles
decay into a baryon and an anti-baryon pair including Roper
resonance to investigate the Roper structure as a hybrid state.

 This paper is organized  as follows: in the second section, we
 present our model of the $\jp $ decay into a hybrid baryon and an
 anti-baryon, and  make a
choice of wave functions for hybrid baryons. With this model
 ,in section 3, we formulate the amplitudes of $\jp$ decays
 into $\bar p\nstar$ and $\bar \nstar \nstar$ baryon pairs ,
 and its corresponding decay widths and
angular distributions for these two decay modes. In the fourth
section,the determination of the quark pair creation strength from
the process $\jp\to \gamma p\bar p$ is formulated. Then, we
present our main numerical results on angular distributions and
decay widths versus the mixing parameter for the two decay modes
$\jp\to \bar{p}N^*,\bar{N}^*N^*$, where the structure of the Roper
resonance in two different pictures are assumed. We also discuss
our results in the last section. Some matrix elements of the
transitional amplitude are appended at the end of paper.
\section{Model description}
 The exclusive decays of the charmonium have been investigated by many
 authors within perturbative QCD\cite{duncan}. The main dynamical mechanism
 is simply assumed that the $c\bar{c}$ quarks annihilate into a minimum number of
 gluons constrained by the charge and parity conservation.
The decay of the $\jp$ particle into a baryon-antibaryon pair is
currently assumed to proceed via two steps as illustrated by
Fig.1(a). In the first step, the $c\bar c$ pair annihilates into
three gluons, followed by the materialization of each gluon into a
pair of quark-antiquarks. Since the $c\bar c$ quarks annihilate
only if their mutual separation is less than about $1/m_c$ ($m_c$
is the mass of c quark), which is smaller than the
non-perturbative charmonium radius, and since the average energy
of three virtual gluons is of the order 1-2 GeV which lies in the
QCD perturbative region. Thus this step is usually carried out
from perturbative QCD to the lowest order of $\alpha_s$. Then in
the second step the three quarks on the one hand and the three
antiquarks on the other hand combine to form a baryon and
antibaryon, respectively. The nonperturbative dynamics is included
by the quark wave functions inside the baryons from the quark
model.

The exclusive decays of the $\jp$ can also provide a new
laboratory to study the Roper properties as a hybrid
state\cite{zou}. Since the strong decay of the $\jp$ particle is a
rich-gluon process, one may indeed expect a possible situation
that a gluon combines with three quarks or antiquarks to form a
physical singlet, a candidate of the hybrid baryon, denoted by
$\st{qqqg}$ as illustrated by Fig.1.(b). In this decay mode, a
quark-antiquark pair with a vacuum quantum number $J^{pc}=0^{++}$
is assumed to create from the QCD vacuum due to the strong
quark-gluon coupling. For the sake of simplicity , the coupling of
the created quark pair is parameterized into an effective strength
$g_I$ within a modified quark pair creation model. Generally
speaking, the strength $g_I$ may be dependent on the model and the
energy level. We feel it is reasonable to determine its value from
the radiative decay of the $\jp$ into $\gamma p\bar p $, which has
an almost identical Feynman diagram as shown in Fig.1(c) as for
the hybrid baryon production in Fig.1(b).
\begin{figure}[htbp]
\begin{center}
\hspace*{-0.cm} \epsfysize=6cm \epsffile{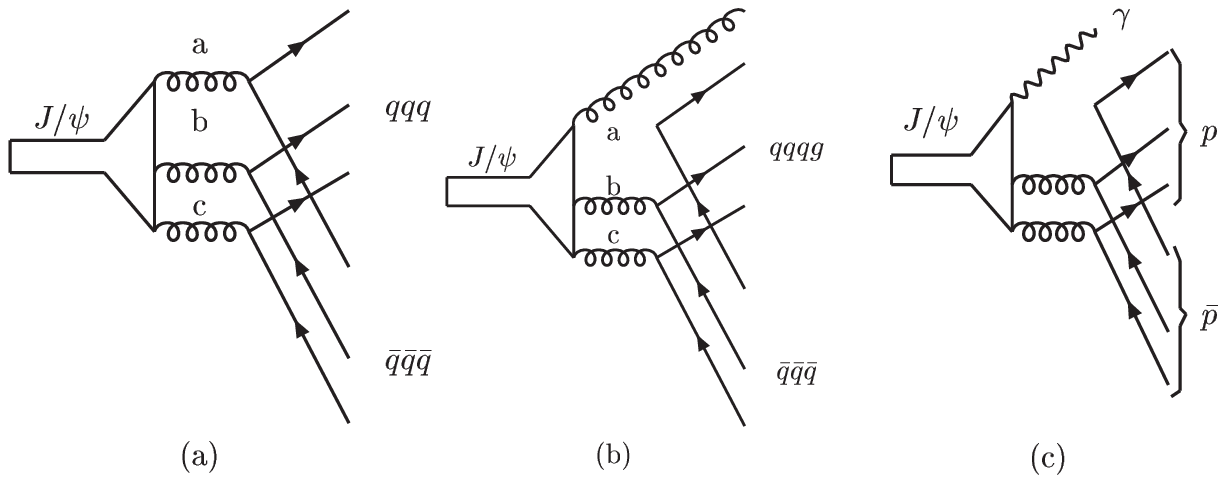}
\end{center}
\caption{Lowest-order diagrams for (a)$\jp\to
\st{\bar{q}\bar{q}\bar{q}}+\st{qqq}$, (b)$\jp\to
\st{\bar{q}\bar{q}\bar{q}}+\st{qqqg}$,(c)$\jp\to \gamma p\bar p$}
\end{figure}

As in naive quark model, the quarks and the gluon bound in hybrid
baryons are treated as constituent components with mass $m_q$ and
$m_g$,respectively, and their spatial wave function are chosen to
be harmonic-oscillator eigen wave functions in the center of mass
(CM) system of the baryon. The construction of the hybrid wave
function is completed by combining the quark and the gluon sector
into a color-singlet state, which is indeed included in the
representation of group $SU(3)_c\bigotimes SU(3)_c\bigotimes
SU(3)_c\bigotimes SU(8)_c$. Let I,J denote the quantum numbers of
total isospin and angular momentum for $\st{qqqg}$ state with
$I_3$ and $J_3$ , the z-projection components, then the wave
function of the hybrid baryon denoted by $\st{Ng,I_3J_3}_{2I,2J}$
can be expressed as follows,

\begin{equation}\label{hwf}
\st{Ng,I_3J_3}_{2I,2J}={1 \over\sqrt{8}}\sum_{\alpha=1}^{8}
\sum_{j_3}C^{J,J_3}_{jj_3,1J_3-j_3}\st{N,I_3j_3}^{\alpha}_{2I,2j}\st{g,1J_3-j_3}^{\alpha}.
\end{equation}
where $\st{N,I_3j_3}^\alpha_{2I,2j}$ is the color-octet $qqq$ part
in the hybrid state, with quantum number of the total isospin $I$
,z-projection component $I_3$ and angular momentum $j$
,z-projection $j_3$. $\alpha$ is the color-index, and $\st{g,1
J_3-j_3}^\alpha$ is the color spin wave function for the gluon.

Especially, for I=1/2, the wave functions for the color-octet
$qqq$ part\cite{barnes} are
\begin{eqnarray}\label{}
^48:\st{N,I_3j_3}^\alpha_{1,3}&=&{1 \over \sqrt{2}}
(\phi^\lambda\psi^\rho_\alpha-\phi^\rho\psi^\lambda_\alpha)\chi^{s},\\
^28:\st{N,I_3j_3}^\alpha_{1,1}&=&{1 \over
2}[(\phi^\rho\chi^\rho-\phi^\lambda\chi^\lambda)\psi^\rho_\alpha-
(\phi^\rho\chi^\lambda+\phi^\lambda\chi^\rho)\psi^\lambda_\alpha)]
.
\end{eqnarray}
Here $\phi,\chi$ and $\psi$ denote the flavor, spin and color wave
functions, respectively. $s,\lambda$ and $\rho$ denote the
symmetry or mixed symmetry representations of the total
permutation of particles ($\lambda$ is symmetric and $\rho$
antisymmetric under (23) interchange). The study of the
electromagnetic transitional properties suggested that the wave
function for the $|qqqg\rangle$ state should be chosen as a mixing
of $^48$ and $^28$ with an equal ratio\cite{li},i.e.
\begin{equation}\label{tmix}
\st{qqqg}={1\over \sqrt
2}(\st{Ng,I_3J_3}_{1,3}+\st{Ng,I_3J_3}_{1,1})\times
\textrm{symmetric~spatial~ part}.
\end{equation}

In the constituent quark model, the wave function of the nucleon
and the Roper resonance are assigned as a ground state and the
first radial excitation of three constituent quarks in the
harmonic oscillator potential, respectively. The nucleon wave
function in the color-spin-flavor space reads,
\begin{equation}\label{nucl}
  \st{p}\equiv\st{qqq,1s}={1 \over \sqrt{2}}\psi^a(\chi^\rho\phi^\rho+\chi^\lambda
\phi ^\lambda)\times \textrm{symmetric~spatial~part},
\end{equation}
where $\psi^a$ is the asymmetric color wave function
,$\chi^\rho(\chi^\lambda)$ and $\phi^\rho(\phi^\lambda)$ are the
spin and flavor wave functions for the $qqq$ cluster with $\rho$
(or $\lambda$)-type mixing symmetry, respectively.

The identification of the Roper resonance as a $\st{qqq,2s}$ state
meets some difficulties in the reproduction of the mass spectra
and the photoproduction amplitudes. On the other hand, some
theoretical studies have predicted the existence of a hybrid state
at this mass region\cite{barnes,martynenko,kisslinger}. So the
mixing of those two configurations should be considered.
Li\cite{li} once mixed the $\st{qqqg}$ part with $\st{qqq,1s}$ in
the study of $N^*(1440)$ in the photoproduction process, where the
wave function of nucleon $|p\rangle$ and the Roper resonance
$|\nstar\rangle$ are assumed to be linear combination of
$|qqq,1s\rangle$ and $|qqqg\rangle$ components. Thus the nucleon
wave function with gluonic freedom degree in the lowest nontrivial
order is expressed by
$|p>=[|qqq,1s\rangle-\xi|qqqg\rangle]/\sqrt{1+\xi^2}$. A low lying
hybrid baryon is assumed to be the Roper resonance, whose wave
function is chosen as to be orthogonal to the nucleon wave
function,i.e.$
|\nstar>=\left[\xi|qqq,1s\rangle+|qqqg\rangle\right]/\sqrt{1+\xi^2}$.
In this scheme ,the wave functions for the nucleon and Roper with
gluonic freedoms  are constructed by introducing a
phenomenological mixing parameter $\xi$ to combine the gluonic
freedom with the conventional proton. Its lowest limit $\xi=0$
corresponds the extreme situation,namely, the nucleon is the
conventional $|3q,1s\rangle$ state, while the Roper is a pure
hybrid state. However,the upper limits of the mixing parameter
$\xi$ is unrestricted. If the fraction of $\st{qqqg}$ part grows
larger in value, it will change the nucleon properties. On the
other hand, there are two states $\st{qqq,2s}$ and $\st{qqqg}$ at
the Roper mass region theoretically. These two states may mix
together due to the residual interaction. One may think that the
mixing of the two states might change their eigen energies. So we
propose an alternative scheme to construct the wave function for
the Roper resonance. It is assigned as a mixture of the
$\st{qqq,2s}$ part with the $\st{qqqg}$ part,
\begin{equation}\label{hybr}
\st{N^*}=\delta\st{qqq,2s}+\sqrt{1-\delta^2}\st{qqqg},
\end{equation}
which is automatically orthogonal to the conventional nucleon
state. Here $\delta$ is a mixing parameter running from 0 up to 1.
If $\delta=0$, it indicates that $\nstar(1440)$ is a pure hybrid
state, and $\delta=1$,the pure $\st{qqq,2s}$ state.

\section{Decay widths and angular distributions}
The full decay width of the $\jp$ particle into three gluons has
been investigated many years ago. Since the charmonium decay
process involves two different physical scale: the binding energy
$\epsilon$ of the bound state and the heavy c-quark mass
$m_c(m_c\gg \epsilon)$, now it is  widely accepted that the
transitional amplitudes for charmonium decays can be expanded to
order $O(p/m)$ , where $p$ is the relative momentum of the
$\ccbar$ quarks. For $\jp$ particles described as a $1s$ bound
state, only the first order term contributes dominantly to the
decay width. This approximation results in that the decay width is
proportional to the value of the $\jp$ wave function at the origin
$|\Phi(0)|$. To study the exclusive decays of $\jp$ into a baryon
anti-baryon pair, we focus on the dynamical behavior of the gluon
properties, the outgoing quarks and the information on the baryon
structure. We feel that the non-perturbative properties of the
bound $\ccbar$ quarks and their dynamical effects do not play
important roles in the $\jp$ exclusive decays at least in the
limit of the non-relativistic approximation, since in which the
contribution from $\ccbar$ quarks are independent on their
relative momenta. Hence the non-perturbative effects within this
energy level, together with the contribution from the $\ccbar$
quarks and the decay constant of $\jp$ and so on are all
parameterized into an overall constant $C_0$.

The gluons from the $\ccbar$ annihilation play a different role in
our model. In the process $\jp\to 3g\to 3\bar q+3q$, each virtual
gluon behaves with equal status due to the fact that each gluon
finally creates a $q\bar q$ quark pair. While in the process
$\jp\to 3g\to 3\bar q+3q+g$, of all three gluons, there being one
gluon does not create a $q\bar q$ quark pair ,and instead is bound
inside the hybrid baryon in the final state. We consider this
gluon to be a constituent one in the bound state.

If we assume the Roper resonance is a pure hybrid state, i.e.
without mixing with $|qqq\rangle$ component in its wave functions,
there only being one precess as shown in fig.1(b) should be
considered. In general, the Roper resonance is assumed to be a
mixture of the $|qqq\rangle$ and $|qqqg\rangle$ components. In
this case of mixing picture, to study the decay of $\jp$ into the
Roper, it is essential to calculate the basic transitional
amplitudes of both decay modes given in fig.1.(a) and (b). The
basic amplitudes for these two processes can be expressed as,
\begin{eqnarray}\label{}
<\bar{N} N|T_1C_1|\jp^{(\Lambda)}>_{s_z,s'_z}&\equiv&
<\Psi_{N}(q',s'_z)\Psi_{\bar N}(q,s)|C_1T_1|\jp^{(\Lambda)}>\nonumber \\
&=&\sum_{s_i,s'_i}\int(\prod^3_{i=1}d\overrightarrow{q}_id\overrightarrow{q}'_i)
<\Psi_{N}(q',s'_z)\Psi_{\bar{N
}}(q,s_z)|q_i,s_i,q'_i,s'_i,i=1..3>\nonumber\\
& \times &<q_i,s_i,q'_i,s'_i,i=1..3|C_1T_1|\jp^{(\Lambda)}>,
\end{eqnarray}
for fig.1.(a) ,and\\
\begin{eqnarray}\label{}
<\bar{N}N_g|T_2C_2|\jp^{(\Lambda)}>_{s_z,s'_z}&\equiv&
<\Psi_h(q',s'_z)\Psi_{\bar{N}}(q,s_z)|C_2T_2|\jp^{(\Lambda)}>\nonumber \\
&=&\sum_{s_i,s'_i}\int(\prod^{j=1..3}_{i=1..4}d\overrightarrow{q}_id\overrightarrow{q}'_j)
<\Psi_h(q',s'_z)\Psi_{\bar{N}}(q,s_z)|q_i,s_i,q'_j,s'_j,i=1..4,j=1..3>\nonumber\\
&\times&<q_i,s_i,q'_j,s'_j,i=1..4,j=1..3|C_2T_2|\jp^{(\Lambda)}>,
\end{eqnarray}
for fig.1.(b). Here $N$ and $N_g$ stand for the $qqq$ and $qqqg$
states, respectively. $\Psi_{N }(q',s'_z)$ and $\Psi_h(q',s'_z)$
are the flavor-spin-spatial wave functions for $qqq$  and $qqqg$
cluster, respectively.

The hard scattering part in above equations can be written
explicitly by the standard Feynman rules,i.e.
\begin{eqnarray} \label{d1}
<q_i,s_i,q'_i,s'_i|C_1T_1|\jp^{(\Lambda)}>&=&C_0<C_1>(ig)^6\epsilon_{\jp}^{(\Lambda)\lambda}{g_{\mu\lambda}g_{\nu\rho}+g_{\nu\lambda}g_{\mu\rho}+g_{\rho\lambda}g_{\mu\nu}
\over
(q_1+q'_1)^2(q_2+q'_2)^2(q_3+q'_3)^2}\nonumber  \\
&\times&\bar{u}(q'_1,s'_1)\gamma^\mu v(q_1,s_1)
\bar{u}(q'_2,s'_2)\gamma^\nu
v(q_2,s_2)\bar{u}(q'_3,s'_3)\gamma^\rho v(q_3,s_3),
\end{eqnarray}
and
\begin{eqnarray}\label{dd2}
<q_i,s_i,q'_j,s'_j|C_2T_2|\jp^{(\Lambda)}>&=&C_0<C_2>(ig)^5g_I\epsilon^{*(\lambda_i)\rho}\epsilon_{\jp}^{(\Lambda)\lambda}{g_{\mu\lambda}g_{\nu\rho}+g_{\nu\lambda}g_{\mu\rho}+g_{\rho\lambda}g_{\mu\nu}
\over
(q_1+q'_1)^2(q_2+q'_2)^2}\nonumber \\
&\times&\bar{u}(q'_1,s'_1)\gamma^\mu v(q_1,s_1)
\bar{u}(q'_2,s'_2)\gamma^\nu
v(q_2,s_2)\bar{u}(q'_3,s'_3)v(q_3,s_3),
\end{eqnarray}
where $\epsilon_{\jp}^{(\Lambda)},\epsilon^{(\lambda_i)\rho}$ are
the polarization vectors for $\jp$ and the gluon with helicity
values $\Lambda$ and $\lambda_i$, respectively. We perform the
calculation in the $\jp$ rest system. For $\jp$ produced in
$e^+e^-$ annihilation, its helicity is limited to be
$\Lambda=\pm1$. $q_i(q'_i)$ and $s_i(s'_i)$ are the four vector
momentum and spin z projection of the anti-quarks (quarks),
respectively. $\langle C_1\rangle$ and $\langle C_2\rangle$ are
color factors corresponding to the decay mode as depicted in
fig.1.(a) and (b), and $g=\sqrt{4\pi\alpha_s}$ is a strong
coupling constant. $g_I$ is an effective strength of the created
quark pairs in the $^3P_0$ quark model which will be determined
from the decay mode $\jp\to \gamma p\bar p$. Other constants and
contributions from charmonium bound properties, which are assumed
to be independent on a given process, are all put into a single
overall constant $C_0$. $u(q'_i,s'_i)$ and $v(q_i,s_i)$ are
assumed to be free Dirac spinors for quarks and anti-quarks,
respectively. They are explicitly given by
\begin{equation}\label{}
u(q'_i,s'_i)=\sqrt{{m_i+E'_i\over 2E'_i}}\Bigg(\matrix{1\cr
{\overrightarrow{\sigma}\cdot\overrightarrow{q}'_i\over
m_i+E'_i}}\Bigg)\chi'_{s_i}, ~~~ v(q_i,s_i)=\sqrt{{m_i+E_i\over
2E_i}}\Bigg(\matrix{
 {\overrightarrow{\sigma}\cdot\overrightarrow{q}_i\over m_i+E_i}\cr 1
 \cr}
 \Bigg)\chi_{s_i},
\end{equation}
where $E_i\equiv q_i^0$ is the energy of a quark.

The evaluation of the color factor $<C_1>$ and $<C_2>$ for these
decay modes is straightforward  in terms of the SU(3) structure
constant $d_{abc}$ and the SU(3) color matrices $T$, they are
\begin{equation}\label{cf1}
<C_1>={1\over
24\sqrt{3}}d_{abc}\epsilon^{ijk}T^a_{il}T^b_{jm}T^c_{kn}\epsilon^{lmn}={5\over
18\sqrt{3}},
\end{equation}

\begin{equation}\label{cf2}
<C_2>={1\over
4\sqrt{18}}d_{abc}\epsilon^{ijk}T^b_{in}T^c_{jm}\delta_{kl}\psi^\rho_a(lmn)={5i\over
144},
\end{equation}
where $T^{(a)}(a=1..8)$ are Gellmann matrices, and
$\psi^\rho(lmn)$ is the color-octet wave function for three quarks
in the $\st{qqqg}$ state with antisymmetry by exchanging quark (2)
and (3), while the $\lambda$-type wave function
$\psi^\lambda(lmn)$, which is symmetric by exchanging quark (2)
and (3), does not have a contribution to color factor $<C_2>$ due
to the symmetric properties of Gellmann matrices.  It is worthy to
note that the ratio $\langle C_2\rangle/\langle C_1\rangle \approx
0.2$ from Eqs. (\ref{cf1}) and (\ref{cf2}), which means that the
decay of the $\jp$ particle into the hybrid state is a color
suppressed process.

The components involving Dirac spinors in Eqs.(\ref{d1},\ref{dd2})
can be
expressed more explicitly as\\
\begin{eqnarray}
\epsilon_{\jp}^{\pm}&=&(0,\mp{1\over \sqrt{2}},-{i\over
\sqrt{2}},0),\nonumber \\
\bar{u}(q'_i,s'_i)\gamma^0v(q_i,s_i)&=&\sqrt{{(E_i+m_i)(E'_i+m_i)\over
4E_iE'_i}}\left <s'_i\left
|{\overrightarrow{\sigma}\cdot\overrightarrow{q}'_i\over m_i+E'_i
}+{\overrightarrow{q_i}\cdot \vec \sigma\over
m_i+E_i}\right|s_i\right >,\nonumber \\
\bar{u}(q'_i,s'_i)v(q_i,s_i)&=&\sqrt{{(E_i+m_i)(E'_i+m_i)\over
4E_iE'_i}}\left <s'_i\left
|{\overrightarrow{\sigma}\cdot\overrightarrow{q}_i\over m_i+E_i
}-{\overrightarrow{q}'_i\cdot \vec \sigma\over
m_i+E'_i}\right|s_i\right >,\nonumber \\
\bar{u}(q'_i,s'_i)\overrightarrow{\gamma}^\mu
v(q_i,s_i)&=&\sqrt{{(E_i+m_i)(E'_i+m_i)\over 4E_iE'_i}}\left
<s'_i\left|\overrightarrow{\sigma}+{(\overrightarrow{\sigma}\cdot\overrightarrow{q'}_i)\overrightarrow{\sigma}(\overrightarrow{\sigma}\cdot\overrightarrow{q}_i)\over
(E_q+m_q)(E'_q+m_q)}\right|s_i\right>,
\end{eqnarray}
 where $E_i$ and $m_i$ are the energy and the
mass of the quark.

From these amplitudes,one may obtain elements of the transitional
amplitudes $M_{s_zs'_z}^{(\Lambda)}$ for processes $\jp\to \bar
pp, \bar p \nstar,\bar N^*\nstar$ by projecting them on the wave
functions of the nucleon or the Roper resonance ( see Appendix B).
In our calculation, we ignore the amplitude for $\jp\to
\st{qqqg}+\st{\bar q\bar q\bar qg}$ because we argue that the
contribution of matrix elements for this process is trivial due to
the color factor and an additionally suppressed factor $e^{-{p^2_c
/2\beta}}$ in the spatial wave function for the $\st{qqqg}$
cluster.
Then the differential decay width  for $\jp\to
B\bar{B}(B:\textrm{baryon})$ can be expressed as,
\begin{equation}\label{}
{d\Gamma(\jp^{(\Lambda)}\to B\bar{B}) \over d\Omega}={1\over
32\pi^2}\{|M^{(\Lambda)}_{{1\over 2}{1\over
2}}|^2+|M^{(\Lambda)}_{-{1\over 2}{1\over
2}}|^2+|M^{(\Lambda)}_{{1\over 2}-{1\over
2}}|^2+|M^{(\Lambda)}_{-{1\over 2}-{1\over
2}}|^2\}{|\overrightarrow{q}|\over M^2_{\jp}},
\end{equation}
where $M_{\jp}$ is the mass for $\jp$, and the
$\overrightarrow{q}$ is the momentum of outgoing baryons.

From the symmetry consideration, the differential decay width of
$\jp\to B\bar{B}$ can be expressed in a more general form,
\begin{equation}\label{}
{d\Gamma(\jp^{(\Lambda)}\to B\bar{B}) \over d\Omega}=N_{
B\bar{B}}(1+\alpha_B cos^2\theta),
\end{equation}
where $N_ {B\bar{B}}$ is a constant directly related to the
experimental total branching ratio of $\jp\to B\bar{B}$, and can
be used to fix the overall constant $C_0$, and $\theta$, is the
angle between the positron beam direction and the direction of the
outgoing baryon. $\alpha_B$ characterizes the angular distribution
of $\jp\to B\bar{B}$, which can be extracted from the experiments.

\section{Strength of quark-pair creation  in $J/\psi$ decays}
The quark-pair creation model was first introduced by
Micu\cite{micu} in 1969 in a study of meson decays, which
suggested that the strong decay of mesons proceeds through a
simple quark-antiquark ($\qqbar$) pair creation from the vacuum,
with a quantum number $J^{pc}=0^{++}$. With a $\qqbar$ pair
creation operator and a dimensionless constants $\gamma$, the
widths of the two-body hadron decays can be evaluated easily from
definite transitional amplitudes. Many authors have developed
Micu's original suggestion and applied it extensively to a number
of baryon and meson decays with considerable successes. In this
model, the constant $\gamma$ is a parameter and is obtained from
the fitting to the light meson or baryon decay widths.

For the OZI rule allowed decays of the $\jp$ particle, the
$\qqbar$ pair might be created from either a gluon annihilation or
the QCD vacuum . However, in the energy region of the $\jp$ decay
the strength of the $\qqbar$ pair creation from the QCD vacuum may
be different from that in the light meson or baryon decays.
Therefore, we adopt the modified quark-pair creation model to
extract the strength of the quark-pair creation by evaluating the
decay width of the process $\jp\to \gamma p\bar p$.

  The radiative decay $\jp\to \gamma p\bar p$ is assumed to proceeds
via two steps as illustrated by fig.1(c). In the first step, the
$c \bar c$ pair annihilates with the emission of a photon and two
gluons. In the second step, each one of the gluons creates a
quark-antiquark pair and another quark-antiquark pair is created
from the vacuum, which can be described by quark creation model
with the quark pair creation strength $g_I$. By a complicated
final state interaction, a proton and an antiproton are formed.
This complicated hadronization process is accomplished by the
bound state quark wave functions of the nucleon, which is
constructed in the constituent quark model.

In our quark pair creation model, the quark-antiquark pair with
any color and flavor can be created anywhere from the QCD vacuum
with equal strength. But only those pairs whose color-flavor wave
function and space wave function overlap with that of baryons in
the final state can make a contribution to the decay width.
Following the usual procedure, the Hamiltonian for the  quark pair
creation can be defined in the modified $^3P_0$ model\cite{Ackleh}
in terms of quark and antiquark creation operators $b^+$ and
$d^+$,
\begin{equation}\label{hamilton}
H_I=\sum_{i,j,\alpha,\beta,s,s'}\int d^3kd^3k'~g_I[\bar
u(\overrightarrow{k}',s')v(\overrightarrow{k},s)]b^+_{\alpha,i}(\overrightarrow{k}'s')d^+_{\beta,j}(\overrightarrow{k},s)\delta
^3(\vec k-\vec k')\delta_{\alpha\beta}\hat C_I,
\end{equation}
where $\alpha (\beta)$ and i(j) are the flavor and color indexes
of the created quarks (anti-quarks) , and $u(k',s')$ and $v(k,s)$
are free Dirac spinors for quarks and antiquarks, respectively.
They are normalized as
$u^+(k,s)u(k,s')=v^+(k,s)v(k,s')=\delta_{ss'}$ .
$\hat{C}_I=\delta_{ij}$ is the color operator for $q\bar q$ and
$g_I$ is the strength of the decay interaction, which will be
extracted from fitting the width $\Gamma(\jp\to \gamma p\bar p)$
to the experimental data.
 In the non-relativistic limit, $g_I$ can be related to $\gamma$,
 the strength of the conventional $^3P_0$
 model, by $g_I = 2m_q\gamma$\cite{Ackleh}.

We first define an operator for $\jp\to \gamma +2q+2\bar q$
transitions,{\it i.e.}

$$\hat{O}(\jp\to \gamma, qq,\bar q\bar q)
=C_0\hat{C}(ie)(ig)^4\epsilon^{(\Lambda)\lambda}_{\jp}
\epsilon^{*(\lambda_i)\rho}{g_{\mu\lambda}g_{\nu\rho}+g_{\nu\lambda}g_{\mu\rho}+g_{\rho\lambda}g_{\mu\nu}
\over (q_1+q'_1)^2(q_2+q'_2)^2}$$
\begin{equation}\label{d2}
\cdot\bar{u}(q'_1,s'_1)\gamma^\mu v(q_1,s_1)
\bar{u}(q'_2,s'_2)\gamma^\nu v(q_2,s_2),
\end{equation}
where $q_i(i=1,2,3)$ and $q'_{i}(i=1,2,3)$ are the four-momentum
vectors for anti-quarks and quarks, respectively. $C_0$ is an
overall constant which can be determined from fitting the decay
width of the process $\jp\to p\bar p$. $\hat{C}={1\over 2\sqrt
3}\delta_{ab}T^a_{kn}T^b_{jm}$ is the color operator in terms of
SU(3) color matrices T, and $g$ is a strong coupling constant.
$\epsilon^{(\lambda_i)}$ is a photon polarization vector with
helicity $\lambda_i=\pm 1$. $\epsilon^{(\Lambda)\lambda}_{\jp}$ is
the $\jp$ polarization vector.

Inside a baryon there are strong interactions among three
constituent quarks. To form a baryon from the three quarks is a
non-perturbative process. Following the usual approach, we account
for all strong interaction effects by the bound-state quark
wave-functions in the final ( anti-)baryon state. The amplitude
for  $\jp\to\gamma p\bar p$ transition can be written as,
\begin{equation}
\mu^{(\Lambda)}(s_z,s'_{z},p_1,p_2,p_3)=\langle\psi_{\bar{p}}(s_{z},p_1,q_1,q_2,q_3)\psi_{p}(s_{z}^{'},p_2,q'_1,q'_2,q'_3)
|\hat{O}H_I|\psi^{(\Lambda)}_{\jp}(P)\rangle,
\end{equation}
where $s_z$ and $s_z'$ are the spin z-projections for the
anti-baryon and baryon, respectively.
$\psi_{\bar{p}}(s_{z},p_1,q_1,q_2,q_3)$ and
$\psi_{p}(s_{z}^{'},p_2,q'_1,q'_2,q'_3)$ are the
spin-flavor-spatial wave functions of the baryon and
anti-baryon,respectively. In general , the structure of
spin-flavor wave functions of baryons can be constructed in the
constituent quark model. The spin and flavor wave functions of the
proton and antiproton are similar, i.e.
\begin{equation}\label{}
\psi^p_{SF}=\psi^{\bar p}_{SF}={1\over \sqrt 2
}(\chi^\rho\phi^\rho+\chi^\lambda\phi^\lambda),
\end{equation}
where $\chi^\rho$ and $\chi^\lambda$ are the spin-${1\over 2}$
wave functions of the quark pair with mixed-symmetry (see
Eqs.(\ref{dr},\ref{dl}) in Appendix A) .

The spatial wave functions for the proton or antiproton are chosen
as simple harmonic-oscillator eigenfunctions in their
center-of-mass(c.m.) system. However, our calculation is carried
out in the $\jp$ rest system, where the outgoing proton and
antiproton are moving with high relativistic speeds. One has to
transfer the quark wave function from the c.m. system of the
baryon to the laboratory system. Here we only perform the Lorentz
boosts for the spatial wave function and ignore the Melosh
rotations of the quark spinors. This approximation will be
discussed in the next section.

 The decay widths for the process $\jp\to\gamma p\bar p$ can be expressed as:
\begin{equation}\label{}
 \Gamma={(2\pi)^4 \over
2M}\overline{\sum_{\Lambda}}\sum_{s_z,s_{z'},\lambda_i}\int|\mu
(\Lambda,s_z,s_{z'},p_1,p_2,p_3)|^2\prod_{i=1}^{3}{d^3\vec
p_i\over (2\pi)^32E_i}\delta^4(P-p_1-p_2-p_3),
\end{equation}
where P and M are the four momentum vector and the mass of the
$\jp$ particle, respectively. In the $\jp$ c.m. system,
$\vec{P}=0$. $E_1,E_2 $ and $ E_3$ are the energies of the
anti-baryon, baryon and photon, respectively.

The photon in the process $\jp\to\gamma p\bar p$ may also come
from the bremsstrahlung of the outgoing proton or anti-proton
following the strong decay $\jp\to p\bar p$. However,this
bremsstrahlung mainly contributes to the low-energy
photons\cite{oset}. On the other hand, the existing measurements
for the radiative decays are performed with the condition $m(p\bar
p)<2.79$GeV/c\cite{Eaton}, which implies that photons with
energies smaller than 130 MeV are cut off. In our calculation we
ignore the contribution from the bremsstrahlung in the final state
to the radiative decay width and perform the calculation
 with the experimental condition to evaluate the quark-pair creation
strength. We use the branching fraction $Br(\jp\to \gamma p\bar
p)=3.8\times 10^{-4}$\cite{PDG},the light quark mass $m_q=0.22GeV$
and the harmonic constant 0.08GeV$^2$ to extract the strength
$g_I=15.40GeV$. We also change the harmonic oscillator parameter
from 0.08GeV$^2$ to 0.16GeV$^2$ and find that the strength $g_I$
is not sensitive to the choice of this parameter.

This value is much larger than that from fitting to the light
meson decay widths by Ackleh {\it et.al.}\cite{Ackleh}. From their
calculation, we get $g_I=7.3\textrm{GeV}$. The difference may
indicate that the strength $g_I$ is energy-dependent.  Besides, in
Ackleh's calculation, the non-relativistic limit $E_q=m_q$ is used
and the Lorentz contraction for the wave functions of outgoing
mesons is ignored. While in our evaluation, the relativistic
transformation on the baryonic radial wavefunctions between two
different systems are carried out.

\section{Numerical results and discussions}
Although the decay of the $\jp$ into three gluons and the creation
of a quark-antiquark pair from one gluon may be treated by
perturbative QCD, the formation of a baryon or hybrid baryon is
certainly beyond the asymptotic regime. We account for this effect
by explicitly including the bound state wave function for the
three-quark cluster or quark-gluon cluster. As in the naive quark
model, we treat the quarks and gluon in the baryon as constituent
components with masses $m_q$ and $m_g$, respectively. We assume
that the momentum distribution of the bound components in the
cluster $\st{qqq}$ and $\st{qqqg}$ can be described by wave
functions $\phi_{nl}(p_\rho,p_\lambda)$ and $\phi_h(p_a,p_b,p_c)$
,respectively. Since the spin-independent potential is used in the
constituent quark model, the spatial wave functions of the
constituents can be chosen as simple harmonic oscillator
eigenfunctions in their center-of-mass (CM) system [see
Eqs.(\ref{hwfp},\ref{hwfh})].

 In the CM system of the $\jp$, the two baryons are moving in opposite
directions with highly relativistic speeds. One has to transfer
the quark wave function from the CM system of the baryon to the
laboratory system. The two baryons' radial wave functions in the
laboratory system are related to the wave functions in their CM
system by the Lorentz transformation \cite{Capstick,Capstick1996}.
In principle the quark spin wave functions have to be changed by
the Melosh rotations of the quark spinors simultaneously. For
simplicity, this relativistic transformation can be simplified by
neglecting Melosh rotation effects in the framework of the naive
quark model as in the studies of the resonance
electroproduction\cite{foster}. Since the non-relativistic
descriptions for constituent quarks are employed in our
calculation, the momentum dependence of the spin wave function can
be ignored. On the other hand, for the $\jp$ decay process we can
take the $z$ axis in the direction of the moving baryons. So the
third component of the quark spin remains the same in both
systems. As in our previous studies on decays of $\jp$ into
baryon-antibaryon pairs in quark model\cite{pingprd}, we found
that this treatment can be compensated by taking the parameters
describing the baryonic properties as effective values from
fitting to experimental data. Therefor, in our present
calculations, we only perform the Lorentz boosts for the spatial
wave function as follows,

\begin{equation}\label{}
\phi_p(\overrightarrow{q'}_\rho,\overrightarrow{q'}_\lambda)=\bigg |{\partial(\overrightarrow{k}_\rho,\overrightarrow{k}_\lambda)\over
\partial(\overrightarrow{q'}_\rho,\overrightarrow{q'}_\lambda)}\bigg
|^{1/2}\phi(\overrightarrow{k}_\rho,\overrightarrow{k}_\lambda),
\end{equation}
\begin{equation}\label{}
\phi_h(\overrightarrow{q}_a,\overrightarrow{q}_b,\overrightarrow{q}_c)=\bigg |{\partial(\overrightarrow{p}_a,\overrightarrow{p}_b,\overrightarrow{p}_c)\over
\partial(\overrightarrow{q}_a,\overrightarrow{q}_b,\overrightarrow{q}_c)}\bigg
|^{1/2}\phi(\overrightarrow{p}_a,\overrightarrow{p}_b,\overrightarrow{p}_c),
\end{equation}

In most quark model calculations, the value of $\alpha $ has been
chosen in the range of $0.06\sim 0.22\textrm{GeV}^2$,which
corresponds to the nucleon radius in the range of
$0.52\sim0.98$fm. The range of $\alpha$ value in our model has
been determined elsewhere by fitting the angular distribution of
$\jp\to p\bar{p}$\cite{pingprd}. We take the mass of the light
quark $m_q=0.22$GeV and $\alpha=0.08\textrm{GeV}^2$,which
corresponds to the central value of the angular distribution
parameter $\alpha_{p\bar{p}}=0.62$ for $\jp\to p\bar{p}$. The
assignment of a mass to the constituent gluon is model-dependent.
For instance, the lattice QCD calculations predict the mass of the
constituent gluon $m_g=0.5\sim0.6$ GeV . But from the
phenomenological theory, the mass of gluons are commonly taken in
the range $m_g=0.3 \sim 0.7$GeV at the level of $\jp$
decays\cite{gmass,consoli}. In our model, if we assume that the
Roper is composed of  one constituent gluon and three light quarks
in $1s$ state, we approximately take the constituent gluon mass as
$m_g\sim M_{\nstar}-M_p\approx 0.5$GeV. In our present
calculation, the gluon is regarded as a relativistic constituent
particle, its mass can also be adjusted to a lower value
$m_g=0.37$GeV as taken in Ref.\cite{ducati}. The coupling constant
is chosen to be $\alpha_s=0.28$\cite{chiang}, and the strength of
the quark pair creation, $g_I$, is determined from the calculation
of the decay width $\Gamma(\jp\to \gamma p\bar p)$ in the previous
part.

\subsection{Roper as a pure hybrid baryon or conventional baryon}
At first, we consider the Roper resonance as a pure hybrid
state,namely a $qqqg$ bound state. With selected parameters
varying within our parameter window, we get the angular
distribution parameters $\alpha_*=0.42\sim
 0.57$ and $\alpha_{**}=(-0.1)-(-0.9)$
for the decay modes $\jp\to \bar p\nstar,\bar \nstar
\nstar$,respectively.

Although the dynamical behavior of gluons and created quarks are
important to the decay mode, we find that the structure of the
bound cluster play a dominant role in the evaluation of
contributions of amplitudes to those decay processes. Varying the
mass of a constituent gluon from 0.37GeV up to 0.50GeV, we find
that the amplitude for the hybrid decay modes is much smaller than
that for the normal decay mode, and one obtains
\begin{eqnarray}\label{}
{|<qqq,1s;qqqg,1s|T_2C_2|\jp^{(\Lambda)}>|_{s_zs'_z}\over
|<qqq,1s;qqq,1s|T_1C_1|\jp^{(\Lambda)}>|_{s_zs'_z}}& <0.15;\\
\nonumber
{|<qqq,2s;qqqg,1s|T_2C_2|\jp^{(\Lambda)}>|_{s_zs'_z}\over
|<qqq,1s;qqq,1s|T_1C_1|\jp^{(\Lambda)}>|_{s_zs'_z}} & <0.04.
\end{eqnarray}
This implies that if the Roper is a pure $qqqg$ hybrid, the decay
widths for the process $\jp\to \bar p\nstar,\bar \nstar \nstar$
are almost less than 2\% and 0.2\% of that for the process $\jp\to
p\bar p$, respectively . Where doses this suppression come from? A
great part is from the depressed color factor due to the different
dynamical behaviors of the quarks and gluons in these two decay
processes. One could imagine the fact that if a gluon to combine
with a color-octet three-quark cluster to form a color singlet
object, not all colored gluons are allowed. This situation is well
understood from the calculation of the color factor $<C_2>$, which
is almost the fifth of the color factor $<C_1>$. This indicates
that $\jp$ decays into a hybrid baryon is a color depressed
process. While in the normal decay of the $\jp$ particle,the
gluons and the constituent quarks appear with an equal status due
to the fact that each of the three gluons decays into a quark and
an anti-quark pair and then three quarks (anti-quarks)form a color
singlet baryon (anti-baryon). The color factor for this process is
much larger. In other words, the decay mode for a hybrid baryonic
production is color depressed process. Other source is coming from
the different behaviors of the constituent quarks and gluons bound
in the baryon. We simply assume that the proton is a pure
three-quark system and the $N^*$ is a pure $qqqg$ system. To
calculate the decay amplitude one has to project the basic
amplitude onto the bound state wave functions of three quarks (and
the gluon). The overlap of the basic amplitude with the wave
functions of three quarks is much larger than that with the wave
functions of three quarks and a gluon. By adjusting the gluon mass
from $m_g=0.37$ GeV up to 0.5GeV, and using the harmonic
oscillator parameter $\alpha=0.08\textrm{GeV}^2$,  we finally
obtain the ratios of the $\jp\to p\bar N^*$ and $\jp\to N^*\bar
N^*$ decay widths to the width of the process $\jp\to p\bar p$,

\begin{eqnarray}
{\Gamma(\jp^{(\Lambda)}\to\bar p\nstar)\over
\Gamma(\jp^{(\Lambda)}\to\bar pp)}\bigg|_{\delta=0} & =& \bigg\{ \begin{array}{ll}1.4\times 10^{-2}&(m_g=0.50\textrm{GeV});\\
 1.8\times 10^{-3}& (m_g=0.37\textrm{GeV});\end{array}\\
{\Gamma(\jp^{(\Lambda)}\to\bar \nstar\nstar)\over
\Gamma(\jp^{(\Lambda)}\to\bar pp)}\bigg|_{\delta=0}& =& \bigg\{ \begin{array}{ll}1.2\times 10^{-4}&(m_g=0.50\textrm{GeV});\\
2.3\times 10^{-4}& (m_g=0.37\textrm{GeV}).\end{array}\nonumber
\end{eqnarray}
We also change the harmonic oscillator parameter of nucleons
$\alpha$ from 0.06$\textrm{GeV}^2$ to $0.21\textrm{GeV}^2$. The
results are plotted in Fig.\ref{widthnn}.

 Another extreme situation  is that  the Roper resonance is a
pure common 3q state, namely $\delta=1$ in Eq.(6), which is
assigned as the first radial excitation state $\st{qqq,2s}$. In
this case, we deal only with one decay mode, $\jp\to
\st{qqq}+\st{\bar q\bar q\bar q}$ as illustrated in fig.1(a). With
the same parameters for the quark masses and  the harmonic
oscillator  the numerical results are
\begin{eqnarray}
{\Gamma(\jp^{(\Lambda)}\to \bar p\nstar)\over
 \Gamma(\jp^{(\Lambda)}\to \bar pp)}\bigg|_{\delta=1}&=&2.0\sim
 4.5,\\
{\Gamma(\jp^{(\Lambda)}\to \bar \nstar\nstar)\over
 \Gamma(\jp^{(\Lambda)}\to \bar pp)}\bigg|_{\delta=1}&=&3.2\sim
 22.0,\nonumber
\end{eqnarray}
 and their angular distribution parameters are
$\alpha_*=0.22\sim
 0.70$ and $\alpha_{**}=0.06\sim 0.08$, respectively.
 This indicates  that
 the the $\jp\to p\bar N^*$  decay width
 is much larger if the $N^*$ is a normal three-quark excitation
 than that if  the $N^*$ is a hybrid.

\subsection{Roper in the hybrid picture mixing with the $|qqq,2s\rangle$
state}
\begin{figure}[htbp]
\begin{center}
\hspace{1cm} \vspace{-0.5cm}
\parbox{0.8\textwidth}{\epsfig{file=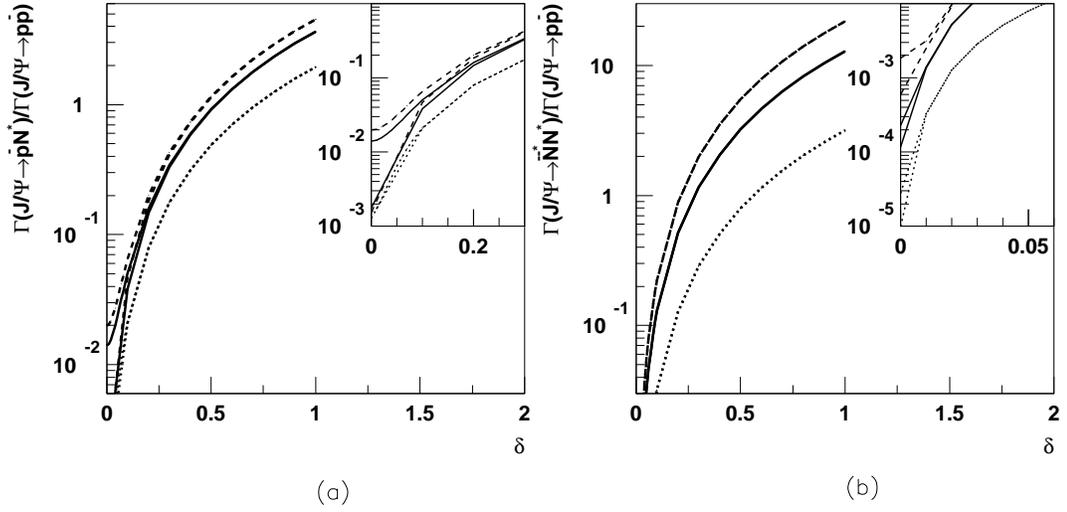,width=0.8\textwidth,silent=,clip=}\\
\vspace{-1cm}\caption{The ratios of decay widths are plotted as a
function of the mixing parameter $\delta$. The $N^* (1440)$ is
assumed as a mixture of the $|qqq,2s\rangle$ and $|qqqg\rangle$
states with mixing parameter $\delta$. The dashed, solid and
dotted curves correspond to the choices of the harmonic oscillator
parameter $\alpha=0.06\textrm{GeV}^2,0.08\textrm{GeV}^2 \textrm{
and } 0.21\textrm{GeV}^2$, respectively. The values for light
quark masses are taken as $m_q=0.22$GeV, (a)~The ratio
$\Gamma(\jp\to \bar{p}N^*)/\Gamma(\jp\to p\bar{p})$.For each type
of curves, the upper one corresponds to the mass of the
constituent gluons $m_g=0.50$GeV and the lower one is related to
the choice $m_g=0.37$GeV . (b)~The ratio $\Gamma(\jp\to
\bar{N}^*N^*)/\Gamma(\jp\to p\bar{p})$.For each type of curves,
the upper solid curve,dash curve and lower dotted curve correspond
to the mass of the constituent gluons $m_g=0.37$GeV and the other
curves are related to the choice $m_g=0.50$GeV . }\label{widthnn}}
\end{center}
\end{figure}

 If the Roper resonance is taken as a mixture of a
normal $|qqq,2s\rangle$ state with a pure hybrid state, the decay
widths for $\jp\to \bar p\nstar,\bar \nstar \nstar$ decay modes
can be evaluated numerically from Eqs.(\ref{pp})$\sim$(\ref{nn}).
The predicted ratios of decay widths in logarithmic scale are
plotted in Fig.\ref{widthnn} in terms of the mixing parameter
$\delta$. The dashed, solid and dotted curves correspond to the
different choice of the harmonic oscillator parameters
$\alpha=0.06\textrm{GeV}^2,0.08\textrm{GeV}^2 $ and
$0.21\textrm{GeV}^2$, respectively. The light quark masses are
taken as $m_q=0.22$GeV. For each type of curves in small $\delta$
region, the upper and lower curves correspond the different choice
of the gluonic mass $m_g=0.37$ or $0.50$GeV as illustrated in the
figure caption. It is clear to see from these results that the
ratios grow rapidly with the increase of the mixing parameter in
small  $\delta$ region. While in larger $\delta$ region, the
ratios almost remain the same as the variation of the gluonic mass
from 0.37GeV up to 0.50GeV. This behavior is easily understood
from calculations of the contributions of the hybrid decay mode to
the transition amplitudes. Since the $\jp$ decays into a hybrid
baryon are color-depressed processes, the normal decay modes
contribute to the transition amplitudes dominantly. Hence
comparing the calculation of the ratios in $\jp$ decays with the
measurements might provide a new approach to probe the structure
of the Roper resonance.

\section{Conclusion}
We have studied the decay widths and angular distributions of the
processes $\jp\to\bar{p}N^*,\bar{N}^*N^*$ in constituent quark
model.  The strong decay of the $\jp$ is assume by the emission of
three gluons. The transitional amplitude of dynamical process for
$\jp\to \st{qqq}+\st{\bar{q}\bar{q}\bar{q}}$ is numerically
evaluated from perturbative QCD approach by assuming that each
gluon creates a quark-antiquark pair. While in the process
$\jp\to\st{\bar{q}\bar{q}\bar{q}}+\st{qqqg}$,we suppose that two
pairs of quark-antiquarks are created from the annihilation of two
gluons and  a light quark pair, $u\bar u$ or $d\bar d$ is created
from the QCD vacuum with strength $g_I$. Then the three quarks (
anti-quarks ) or three quarks and  a gluon are respectively
combine together to form an ordinary baryon or a hybrid state by
some final-state interactions.  The determination of the strength
$g_I$ is also carried out with the modified quark creation model
by fitting the decay width for the process $\jp\to \gamma p\bar p$
to the experimental value, and the numerical result gives the
strength $g_I=15.40$ GeV by using the same experimental cut-off
condition on the restriction of the outgoing photon energy. Then
the transitional amplitudes and angular distribution parameters
for the processes $\jp\to \bar p \nstar,\bar \nstar\nstar$ are
calculated numerically. If the Roper resonance is assigned as a
pure hybrid state, our numerical results show that the ratio $
\Gamma(\jp^{(\Lambda)}\to\bar p\nstar)/
\Gamma(\jp^{(\Lambda)}\to\bar pp)  <2\%, \textrm{and }
\Gamma(\jp^{(\Lambda)}\to\bar \nstar\nstar)/
\Gamma(\jp^{(\Lambda)}\to\bar pp)  <0.2\%$, and their angular
distribution parameters are $\alpha_*=0.42\sim
 0.57$ and $\alpha_{**}=(-0.1)-(-0.9)$, respectively.
However, when the Roper resonance is assumed to be a common $2S$
state, the results are quite different, with $
\Gamma(\jp^{(\Lambda)}\to\bar p\nstar)/ \Gamma(\jp^{(\Lambda)}\to
\bar pp)  = 2.0\sim 4.5, \textrm{and }
\Gamma(\jp^{(\Lambda)}\to\bar \nstar\nstar)/
\Gamma(\jp^{(\Lambda)}\to\bar pp)  = 3.2\sim 22.0 $, and with the
angular distribution parameter $\alpha_*=0.22\sim 0.70
,\alpha_{**}=0.06\sim 0.08$. This implies
 that, not only the dynamics of three gluons and created
quarks , but also the structure of the final cluster state , i.e.
$\st{qqq}$ or $\st{qqqg}$, play important roles in the evaluation
of the amplitudes in these decay processes. So it is suggestive
that an accurate measurement of the decay widths and angular
distributions of these channels may provide us a novel tool to
probe the structure of the Roper resonance. In our present
calculation we also consider the mixing of the normal Roper state
with the hybrid state. Under this picture, we calculate the decay
widths for these decays versus the mixing parameter $\delta$.

In conclusion, we find that the contributions from the Roper
structure to the transitional amplitudes paly an important role in
evaluating the decay widths and angular distribution parameters
for processes $\jp\to\bar{p}N^*$ or $\bar{N}^*N^*$. Hence, the
study of the transitional amplitudes for these decay modes may
provide us a new laboratory to justify the nature of the Roper
resonance. If an accurate experimental data on decay widths and
angular distribution parameters for these processes are available
in the future, one may discriminate the hybrid Roper baryons from
conventional baryons.

This work is supported in part by the National Natural Science
Foundation of China under grand Nos. 10225525, 10055003 and by the
Chinese Academy of Sciences under project No. KJCX2-SW-N02. We
thank E.Oset,J.H\"{u}efner,H.J.Pirner,Z.Y.Zhang, P.N.Shen and
J.G.Bian for useful discussions and comments.
\newpage
{\center{\bf APPENDIX}\\}
\appendix
\section{The construction of the hybrid state wave function}
Let I, J denotes the quantum number of total isospin and momentum
for $|qqqg\rangle$ state with $I_3$ and $J_3$, the z-projection
component,and the wave function of hybrid state denoted by $|N
g,I_3J_3\rangle_{2I2J}$ can be constructed according
Eq.(\ref{hwf}) by combining the gluonic part and the three quarks
part's flavor-spin-color wave function into a color singlet state.
Here, we use the dummy index sum as ${1\over
\sqrt{8}}\sum_{\alpha=1..8}$. Especially, for I=J=1/2,we give
explicitly their form in spin-flavor-color space as follows:
\begin{eqnarray}
\str{N g,{1 \over 2}{1 \over 2}}_{13}&=&{1 \over\sqrt{2}}\str{-1}^{\alpha}{1 \over\sqrt{2}}(\phi^{\lambda}\psi^{\rho}_{\alpha}-\phi^{\rho}\psi^{\lambda}_{\alpha})\chi^{s}_{3/2}\nonumber \\
&-&\sqrt{1\over 3}\str{0}^{\alpha}{1 \over\sqrt{2}}(\phi^{\lambda}\psi^{\rho}_{\alpha}-\phi^{\rho}\psi^{\lambda}_{\alpha})\chi^{s}_{1/2} \nonumber \\
&+&\sqrt{1 \over 6}\str{1}^{\alpha}{1
\over\sqrt{2}}(\phi^{\lambda}\psi^{\rho}_{\alpha}-\phi^{\rho}\psi^{\lambda}_{\alpha})\chi^{s}_{-1/2},
\end{eqnarray}
\begin{eqnarray}
\str{N g,{1 \over 2}-{1 \over 2}}_{13}&=&{1 \over\sqrt{2}}\str{1}^{\alpha}{1 \over\sqrt{2}}(\phi^{\lambda}\psi^{\rho}_{\alpha}-\phi^{\rho}\psi^{\lambda}_{\alpha})\chi^{s}_{-3/2}\nonumber \\
&-&\sqrt{1\over 3}\str{0}^{\alpha}{1 \over\sqrt{2}}(\phi^{\lambda}\psi^{\rho}_{\alpha}-\phi^{\rho}\psi^{\lambda}_{\alpha})\chi^{s}_{-1/2}\nonumber \\
&+&\sqrt{1 \over 6}\str{-1}^{\alpha}{1
\over\sqrt{2}}(\phi^{\lambda}\psi^{\rho}_{\alpha}-\phi^{\rho}\psi^{\lambda}_{\alpha})\chi^{s}_{1/2},
\end{eqnarray}
\begin{eqnarray}
\str{N g,{1 \over 2}{1 \over 2}}_{11}&=&\sqrt{2\over
3}\str{1}^{\alpha}{1\over
2}[(\phi^{\rho}\chi^{\rho}_{-1/2}-\phi^{\lambda}\chi^{\lambda}_{-1/2})\psi^{\rho}_{\alpha}
-(\phi^{\rho}\chi^{\lambda}_{-1/2}+\phi^{\lambda}\chi^{\rho}_{-1/2})\psi^{\lambda}_{\alpha}]\nonumber \\
&-&\sqrt{1\over 3}\str{0}^{\alpha}{1\over
2}[(\phi^{\rho}\chi^{\rho}_{1/2}-\phi^{\lambda}\chi^{\lambda}_{1/2})\psi^{\rho}_{\alpha}-(\phi^{\rho}\chi^{\lambda}_{1/2}+\phi^{\lambda}\chi^{\rho}_{1/2})\psi^{\lambda}_{\alpha}],
\end{eqnarray}
\begin{eqnarray}
\str{N g,{1 \over 2}-{1 \over 2}}_{11}&=&\sqrt{1\over
3}\str{0}^{\alpha}{1\over
2}[(\phi^{\rho}\chi^{\rho}_{-1/2}-\phi^{\lambda}\chi^{\lambda}_{-1/2})\psi^{\rho}_{\alpha}-(\phi^{\rho}\chi^{\lambda}_{-1/2}+\phi^{\lambda}\chi^{\rho}_{-1/2})\psi^{\lambda}_{\alpha}]\nonumber \\
&-&\sqrt{2\over 3}\str{-1}^{\alpha}{1\over
2}[(\phi^{\rho}\chi^{\rho}_{1/2}-\phi^{\lambda}\chi^{\lambda}_{1/2})\psi^{\rho}_{\alpha}-(\phi^{\rho}\chi^{\lambda}_{1/2}+\phi^{\lambda}\chi^{\rho}_{1/2})\psi^{\lambda}_{\alpha}],
\end{eqnarray}
where $\phi,\chi,\psi$ are respectively responsible for the flavor
,spin,and color wave function of the three quark parts in
$|qqqg\rangle$ cluster. The index $s,\lambda,\rho$ denote total
permutation symmetry or mixed representation. The mixed symmetric
states are define as:
\begin{equation}\label{dr}
\rho~\textrm{type}:~\st{xyz}^{\rho}={1\over
2}(\st{xyz}-\st{xzy}+\st{yxz}-\st{yzx}),
\end{equation}
\begin{equation}\label{dl}
\lambda~\textrm{type}:~\st{xyz}^{\lambda}={1\over
2\sqrt{3}}(\st{xyz}+\st{xzy}+\st{yxz}+\st{yzx}-2\st{zxy}-2\st{zyx}),
\end{equation}
if x=y, we must replace $\st{xyz}+\st{yxz}$ with
$\sqrt{2}\st{xxz}$ in the above equation. For example,the mixed
symmetric spin wave functions read:
\begin{equation}
\chi^{\rho}_{1/2,1/2}={1\over
\sqrt{2}}(\st{\uparrow\uparrow\downarrow}-\st{\uparrow\downarrow\uparrow}),
\end{equation}
\begin{equation}
\chi^{\lambda}_{1/2,1/2}={1\over
\sqrt{6}}(\st{\uparrow\uparrow\downarrow}+\st{\uparrow\downarrow\uparrow}-2\st{\downarrow\uparrow\uparrow}),etc.
\end{equation}
\section{The matrix elements of the amplitude for $\jp\to \st{\bar q\bar q\bar q}+\st{qqqg}$}
Since the hybrid wave function is chosen as a mixture of the $^48$
state with the $^28$ state with equal ratio,e.g.
Eq.(\ref{tmix}),we should calculate the matrix elements $<N
|C_2T_2|N g,I_3J_3>_{13}$ and $<N |C_2T_2|N g,I_3J_3>_{11}$ for
the process $\jp\to\st{\bar q\bar q\bar q }+\st{qqqg}$. Note that
$<\psi^a|C_2|\psi^\lambda>=0$, where $\psi^a$ is the asymmetric
color wave function of baryon , so we only calculate the part
$<\psi^a|C_2|\psi^\rho>$.
\begin{eqnarray}
<N |C_2T_2|N g,{1 \over 2}{1 \over 2}>_{13}&=&{1 \over
2\sqrt{2}}|-1>^{\alpha}\elmnt{\psi^{a}}{C}{\psi^{\rho}_{\alpha}}\elmnt{\chi^{\lambda}}{T}{\chi^{s}_{3\over2}} \nonumber \\
 & &-{1
\over\sqrt{12}}|0>^{\alpha}\elmnt{\psi^{a}}{C}{\psi^{\rho}_{\alpha}}\elmnt{\chi^{\lambda}}{T}{\chi^{s}_{1\over2}}
\nonumber \\
& &+{1
\over\sqrt{24}}|1>^{\alpha}\elmnt{\psi^{a}}{C}{\psi^{\rho}_{\alpha}}\elmnt{\chi^{\lambda}}{T}{\chi^{s}_{-
{1\over2}}}
\end{eqnarray}
\begin{eqnarray}
<N |C_2T_2|N g,{1 \over 2}-{1 \over 2}>_{13}&=& {1 \over
2\sqrt{2}}|1>^{\alpha}\elmnt{\psi^{a}}{C}{\psi^{\rho}_{\alpha}}\elmnt{\chi^{\lambda}}{T}{\chi^{s}_{-{3\over2}}}
\nonumber \\
& &-{1
\over\sqrt{12}}|0>^{\alpha}\elmnt{\psi^{a}}{C}{\psi^{\rho}_{\alpha}}\elmnt{\chi^{\lambda}}{T}{\chi^{s}_{-{1\over2}}}
\nonumber \\
& &+{1
\over\sqrt{24}}|-1>^{\alpha}\elmnt{\psi^{a}}{C}{\psi^{\rho}_{\alpha}}\elmnt{\chi^{\lambda}}{T}{\chi^{s}_{{1\over2}}}
\end{eqnarray}
\begin{eqnarray}
<N |C_2 T|N g,{1 \over 2}{1 \over 2}>_{11}&=& {1 \over
2\sqrt{3}}|1>^{\alpha}[\elmnt{\psi^{a}}{C}{\psi^{\rho}_{\alpha}}\elmnt{\chi^{\rho}}{T}{\chi^{\rho}_{-{1
\over
2}}}-\elmnt{\psi^{a}}{C}{\psi^{\rho}_{\alpha}}\elmnt{\chi^{\lambda}}{T}{\chi^{\lambda}_{-{1
\over 2}}}]\nonumber \\
& &-{1 \over
2\sqrt{6}}\str{0}^{\alpha}[\elmnt{\psi^{a}}{C}{\psi^{\rho}_{\alpha}}\elmnt{\chi^{\rho}}{T}{\chi^{\rho}_{{1
\over 2}}}
-\elmnt{\psi^{a}}{C}{\psi^{\rho}_{\alpha}}\elmnt{\chi^{\lambda}}{T}{\chi^{\lambda}_{{1
\over 2}}}]\nonumber \\
\end{eqnarray}
\begin{eqnarray}
<N |C_2T_2|N g,{1 \over 2}-{1 \over 2}>_{11}&=& -{1 \over
2\sqrt{3}}|-1>^{\alpha}[\elmnt{\psi^{a}}{C}{\psi^{\rho}_{\alpha}}\elmnt{\chi^{\rho}}{T}{\chi^{\rho}_{{1
\over
2}}}-\elmnt{\psi^{a}}{C}{\psi^{\rho}_{\alpha}}\elmnt{\chi^{\lambda}}{T}{\chi^{\lambda}_{{1
\over 2}}}]\nonumber \\
& &+{1 \over
2\sqrt{6}}\str{0}^{\alpha}[\elmnt{\psi^{a}}{C}{\psi^{\rho}_{\alpha}}\elmnt{\chi^{\rho}}{T}{\chi^{\rho}_{-{1
\over 2}}}
-\elmnt{\psi^{a}}{C}{\psi^{\rho}_{\alpha}}\elmnt{\chi^{\lambda}}{T}{\chi^{\lambda}_{-{1
\over 2}}}]
\end{eqnarray}
From those elements,one obtains the transitional amplitudes
$M_{s_zs'_z}^{(\Lambda)}$for processes $\jp\to \bar p \nstar,\bar
N^*\nstar$ by projecting them onto the wave functions of the
proton or the Roper resonance.We ignore the trivial contribution
from the process $\jp\to |qqqg\rangle+|\bar q\bar q\bar q
g\rangle$ to the transitional amplitude. if one assumes that Roper
is the mixture of $|qqq,2s>$ state with the hybrid state $|qqqg>$,
one gets
\begin{equation}\label{pp}
M^{(\Lambda)}_{s_z,s'_z}(\jp^{(\Lambda)}\to \bar pp)=<\bar q\bar
q\bar q,1s;qqq,1s|T_1C_1|\jp^{(\Lambda)}>,
\end{equation}
\begin{equation}\label{pn}
M^{(\Lambda)}_{s_z,s'_z}(\jp^{(\Lambda)}\to \bar p \nstar)=\delta
<\bar q\bar q\bar
q,1s;qqq,2s|T_1C_1|\jp^{(\Lambda)}>+{\sqrt{1-\delta^2}} <\bar
q\bar q\bar q,1s;qqqg|T_2C_2|\jp^{(\Lambda)}>,
\end{equation}
\begin{equation}\label{nn}
M^{(\Lambda)}_{s_z,s'_z}(\jp^{(\Lambda)}\to \bar N^*
\nstar)=\delta <\bar q\bar q\bar
q,2s;qqq,2s|T_1C_1|\jp^{(\Lambda)}>+{\sqrt{1-\delta^2}} <\bar
q\bar q\bar q,2s;qqqg|T_2C_2|\jp^{(\Lambda)}>.
\end{equation}
In the decay mode $\jp\to \bar \nstar \nstar$, we only consider
the process illustrated in Fig.1(b), where the normal Roper state
is treated as a $|qqq,2s>$ state, while the hybrid state is
assumed to be a mixture of a $|qqq,2s>$ state and a $|qqqg,1s>$
state.
\section{The radial wave function for $|qqq\rangle$ and $|qqqg\rangle$ clusters}
The radial wave functions of protons or hybrid baryons are chosen
as simple harmonic oscillator eigenfunctions in their
center-of-mass (CM) system,i.e
\begin{equation}\label{hwfp}
\phi_{1s}(\overrightarrow{k_\rho},\overrightarrow{k_\lambda})={1\over(\pi\alpha)^{3/2}}e^{-{(\overrightarrow{k_\rho}^2+\overrightarrow{k_\lambda}^2)}\over
2\alpha},
\end{equation}
for $\st{qqq,1s}$ cluster,and
\begin{equation}\label{hwfh}
\phi_{2s}(\overrightarrow{k_\rho},\overrightarrow{k_\lambda})=\sqrt
3 {1\over(\pi\alpha)^{3/2}}(1-{1\over 3\alpha
}(\overrightarrow{k_\rho}^2+\overrightarrow{k_\lambda}^2))e^{-{(\overrightarrow{k_\rho}^2+\overrightarrow{k_\lambda}^2)}\over
2\alpha},
\end{equation}
for $\st{qqq,2s}$ cluster,where
$\alpha=(3km_q)^{1/2}$,$\overrightarrow{k}_\rho,\overrightarrow{k}_\lambda$
are relative coordinates defined as
\begin{eqnarray}
\overrightarrow{k}_\rho & = &{1\over
\sqrt{6}}(\overrightarrow{k}_1+\overrightarrow{k}_2-2\overrightarrow{k}_3),\\
\overrightarrow{k}_\lambda & = & {1\over
\sqrt{2}}(\overrightarrow{k}_1-\overrightarrow{k}_2).
\end{eqnarray}
Let $p_1,p_2,p_3$ be momenta for three quarks and $p_4$ for a
gluon in the cluster $qqqg$ as depicted in Fig.3, and we define
three relative coordinates as
\begin{eqnarray}
\vec p_a & = &{1\over \sqrt{2}} (\vec p_1-\vec p_2),\\
\vec p_b & = &{1\over \sqrt{6}} (\vec p_1+\vec p_2-2\vec p_3),\\
\vec p_c & = &{1\over \sqrt{12}} (\vec p_1+\vec p_2+\vec p_3-3\vec
p_4).
\end{eqnarray}
One has a spatial wave function for $qqqg$ cluster in momentum
space.
\begin{equation}\label{}
\phi_h(\overrightarrow{p_a},\overrightarrow{p_b},\overrightarrow{p_c})={1\over(\pi\alpha)^{3/2}}e^{-{(\overrightarrow{p_a}^2+\overrightarrow{p_b}^2)}\over
2\alpha}{1\over (\pi\beta)^{3/4}}e^{-{\overrightarrow{p}_c^2 \over
2\beta}},
\end{equation}
where $\beta=(3k\widetilde{M}_g)^{1/2}$ is harmonic-oscillator
parameter ($\widetilde{M}_g={2m_qm_g\over m_q+m_g}$).
\begin{figure}[htbp]
\vspace*{-0.0cm}
\begin{center}
\hspace*{-0.cm} \epsfysize=6cm \epsffile{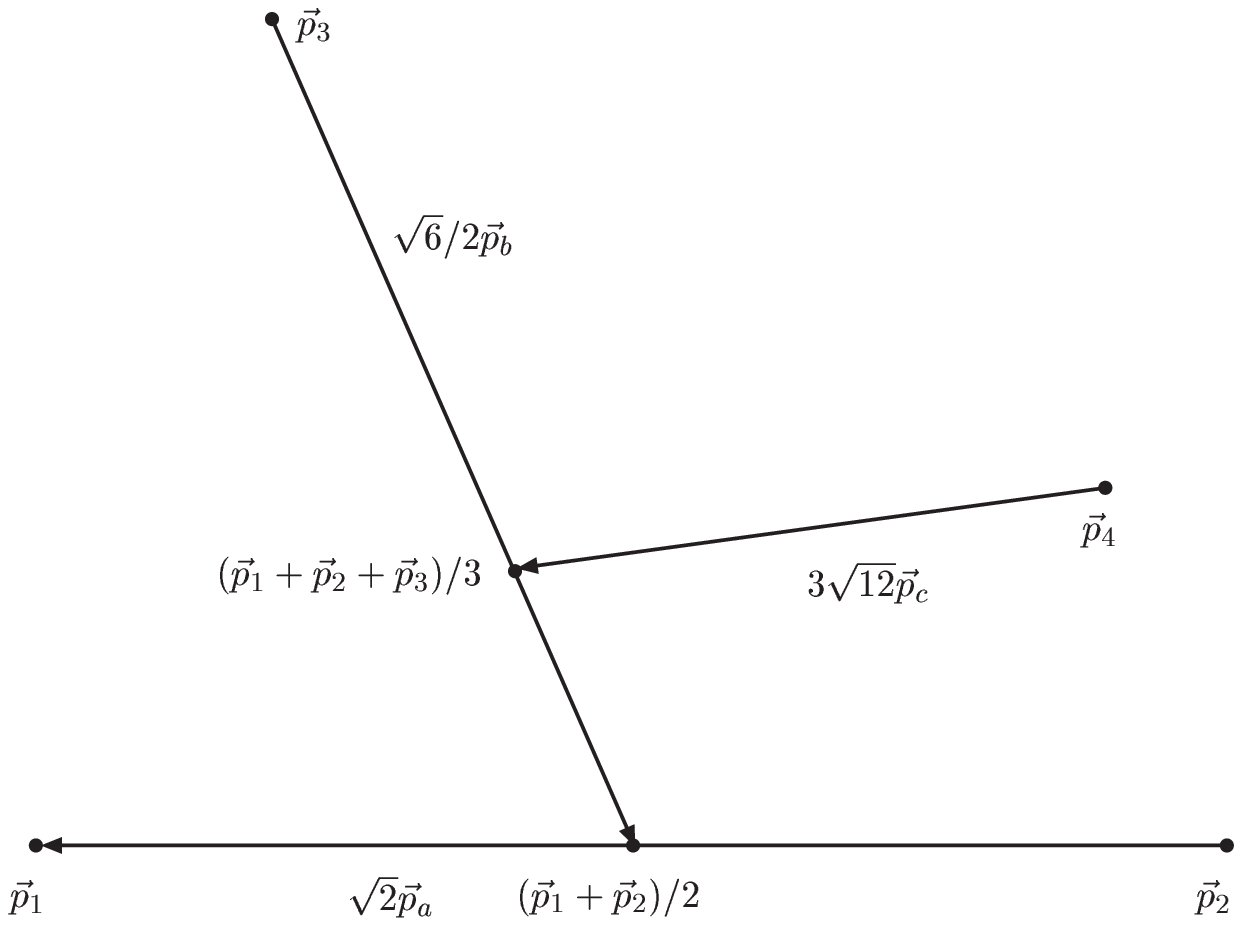}
\end{center}
\vspace*{-0.0cm} \caption{The definition of the relative momentum
for four bodies system in CM system}
\end{figure}


\newpage
\begin{thebibliography}{99}
\bibitem{qm}
M.Gell-Mann,Phys.Rev.125(1962),1067;\\
Y.Ne'eman, Nucl.Phys.26(1961),222.
\bibitem{capstick&r} Simon Capstick and W.
Roberts,Prog.Part.Nucl.Phys.45(2000),S241.
\bibitem {senba}
K.Senba,M.Tanimoto,
Phys.Rev.D28(1983) 2786;\\
L.Kopke,N.Wermes,Phy.Rep.174(1989) 67;\\
Z.Li,V.Burker,Z.J.Li, Phy.Rev.D46(1992) 70.
\bibitem{pi1400}
G.M.Beladidze, et al.,Phys.Lett.B313(1993),276;\\
D.V.Amelin et al.,Phys.Lett.B356(1995),595;\\
S.U.Chung et al.,Phys.Rev.D60(1999),092001;\\
A.Abele, et,al,Phys.Lett.B423(1998),175;\\
T.Barnes,hep-ph/0007296.
\bibitem{pi1600}
V.Dorofeev,VES Collaboration,hep-ex/995002;G.S.Adams,et
al.Phys.Rev.Lett.81(1998),5760.
\bibitem {barnes}
T.Barnes,F.E.Close, Phys.Lett.123B(1983) 89.
\bibitem{golowich}
E.Golowich,E.Haqq,G.Karl, Phys.Rev.D28(1983) 160.
\bibitem{martynenko}
A.P.Martynenko,Sov.J.Nucl.Phys.54(1991) 488.
\bibitem{kisslinger}
L.S.Kisslinger,Z.Li,Phys.Rev.D51(1995) R5986.
\bibitem{Jdecays}
F.E.Close,G.R.Farrar and Z.Li,Phys.Rev.D55(1997),5749;\\
Xue-Qian Li and P.R.Page, Eur.Phys.J.C1(1998),579.
\bibitem{li}
Z.Li, Phys.Rev.D44(1991) 2841.
\bibitem{zou}
B.S.Zou, G.X.Peng, R.G.Ping,et.al.,Eur.Phys.J. A11(2001) 341;\\
B.S. Zou, H.B.Li,et.al.,hep-ph/0004220.
\bibitem{duncan}
A.Duncan,A.Mueller,Phys.Lett.B93(1980) 119;\\
S.J.Brodsky and G.P.Lepage,Phys.Rev.D24(1981) 2848;\\
V.L.Chernyak and I.R.Zhitnitsky,Nucl.Phys.B246(1984) 52;\\
J. Bolz and P. Kroll,Eur.Phys.J.C2(1998) 545.
\bibitem{chiang}
H.C.Chiang,et.al. Phys.Lett.B{\bf324}(1994) 428.
\bibitem{micu}
L.Micu, Nucl.Phys.B10(1969) 521.
\bibitem{Ackleh}
E.S.Ackleh,T.Barnes and E.S.Swanson Phys. Rev.D54(1996) 6811.
\bibitem{oset}
J.A.Oller ,E.Oset and A.Romas  Prog.in Part. and Nucl.
Phys.45(2000) 157.
\bibitem{Eaton}
M.W.Eaton et.al. Phys.Rev.D 29(1984) 804.
\bibitem{PDG}
Particle Data Group, Phys.Rev.D 66(2002) 010001-722.
\bibitem{Capstick}
S.Capstick,B.D.Keister, Phys. Rev. D51(1995) 3598 .
\bibitem{Capstick1996}
S.Capstick,B.D.Keister,27 Nov.199,/nucl-th/9611055 6.
\bibitem{foster}
F.Foster and G.Hughes , Zeitschrift fur physik C14(1982) 123;\\
Q.Zhao,B. Saghai ,et.al.J.Phys.G28(2002) 1293 .
\bibitem{pingprd}
R.G.Ping, H.C.Chiang, B.S.Zou, Phys.Rev.D66(2002),054020.
\bibitem{gmass}
A. C. Aguilar, A. Mihara, and A. A. Natale, Phys.Rev.D65(2002),
054011;\\
Wei-Shu Hou and Gwo-Guang Wong, Phys.Rev.D67(2003),034003;\\
A. Mihara and  A.A. Natale,Phys.lett.B482(2000), 378.
\bibitem{ducati}
M.B.Gay Ducati,F.Halzen, and A.A.Natale, Phys.Rev.D48(1993) 2324.
\bibitem{consoli}
M.Consoli and J.H.Field, Phys.Rev.D49(1994) 1293;\\
M.Consoli and J.H.Field, J.Phys.G23(1997) 41.

\end{thebibliography}
\end{document}